\documentclass[usegraphicx,useAMS,usenatbib]{mn2e}

\newcommand{\mnras}{MNRAS}
\newcommand{\apj}{ApJ}
\newcommand{\apjs}{ApJS}
\newcommand{\procspie}{Proc.\ SPIE}

\newcommand{\mum}{$\umu$m}

\begin{document}
\title[Automated processing of SCUBA data]
{Towards the automated reduction and calibration of SCUBA data
from the James Clerk Maxwell Telescope}

\author[T.\ Jenness et al.]
{T.\ Jenness,$^1$  J.\ A.\ Stevens,$^{2,3}$  E.\ N.\ Archibald,$^1$ F.\
Economou,$^1$ N.\ E.\ Jessop,$^1$
\newauthor
E.\ I.\ Robson$^{1,4}$
\\
$^1$ Joint Astronomy Centre, 660 N. A`oh\={o}k\={u} Place, University Park,
Hilo, Hawaii, 96720, USA \\
$^2$ Mullard Space Science Laboratory, University College London, Holmbury
St. Mary, Dorking, Surrey, RH5 6NT \\
$^3$ Astronomy Technology Centre, Royal Observatory, Blackford Hill, Edinburgh,
EH9 3HJ\\
$^4$ Centre for Astrophysics, University of Central Lancashire, Preston, PR1
2HE
}
\date{Accepted 2002 April 24. Received 2002 March 27; in original form 2001 July 24}

\maketitle
\begin{abstract}
  The Submillimetre Common User Bolometer Array (SCUBA) instrument has been
  operating on the James Clerk Maxwell Telescope (JCMT) since 1997. The data
  archive is now sufficiently large that it can be used to investigate
  instrumental properties and the variability of astronomical sources.  This
  paper describes the automated calibration and reduction scheme used to
  process the archive data with particular emphasis on `jiggle-map'
  observations of compact sources. We demonstrate the validity of our
  automated approach at both 850- and 450-\mum\ and apply it to several of
  the JCMT secondary flux calibrators.  We determine light curves for the
  variable sources IRC+10216 and OH~231.8. This automation is made possible by
  using the ORAC-DR data reduction pipeline, a flexible and extensible data
  reduction pipeline that is used on UKIRT and the JCMT.
\end{abstract}


\section{INTRODUCTION}

The Submillimetre Common User Bolometer Array (SCUBA)
\citep{1999MNRAS.303..659H} consists of two arrays of bolometers (or pixels);
the Long Wave (LW) array has 37 pixels operating in the 750- and 850-\mum\ 
atmospheric transmission windows, while the Short Wave (SW) array has 91
pixels for observations at 350- and 450-\mum. Each of the pixels has
diffraction-limited resolution on the telescope (approximately 14.5 and 8
arcsec FWHM respectively), and are arranged in a close-packed hexagon.  Both
arrays have approximately the same field-of-view on the sky (diameter of 2.3
arcmin), and can be used simultaneously by means of a dichroic.

SCUBA was commissioned in late 1996 and there now exists a large searchable
archive of observations\footnote{The public archive is hosted by the Canadian
Astronomy Data Center}. As of February 2001 the
archive contained approximately 80,000 separate observations.
These data provide an excellent historical record
of the telescope and instrument behaviour as well as being an archive of
astronomical data.  The major challenge of an archive of this size is the data
processing. In order to extract useable historical data from the archive it is
not feasible to reduce the data manually and an automated approach must be
used.

There are three reasons why automated use of the archive is required.  Firstly,
it provides a unique resource for the investigation of instrument and telescope
parameters, leading ultimately to an optimisation of observing practices,
secondly the proposed Virtual Observatories will require reduced and
calibrated data \citep[e.g.][]{2002ADASS}, and
thirdly it allows monitoring of variable sources, especially extragalactic
radio sources that are observed frequently for pointing purposes. In this
latter respect the present paper forms a companion to our recent work on this
topic \citep*{RSJ2001}.

In order to determine its accuracy, any automated technique under consideration
must be used to process a standard calibration source of known properties.  We
present our choice of such a reduction method below, using Uranus as the
primary calibrator and then apply the technique to Mars and several compact
secondary calibration sources as a consistency check.

\section{OBSERVATIONS}

In order to test any automated reduction system it is important that a
suitable dataset is extracted from the archive. This dataset should contain
frequent observations of a bright, compact source of known flux taken with a
standard observing mode.  Although Mars is the primary calibrator at
submillimetre wavelengths, Uranus is a better choice for our approach because
of its compactness relative to the JCMT beam (semi-diameter approximately 1.7
arcsec over the period in question). The best sampled, homogeneous dataset
comes from jiggle-map mode \citep[e.g.,][]{1999MNRAS.303..659H} observations of
Uranus that are performed at 850/450-\mum\ for calibration purposes, and at
850-\mum\ for pointing checks.  We have searched the archive for all
jiggle-map observations of Uranus; between April 1997 (the date when the SCUBA
data headers were finalised) and May 2001 there were approximately 1500
850-\mum\ observations that were performed in conditions suitable for
accurate extinction correction, with 300 of these including data at 450-\mum.
These data use chop throws of between 60 (all the pointing observations) and
120 arcsec (most of the maps).  In October 1999 both the 450- and 850-\mum\
filters were updated to new wideband versions (denoted with a `W' suffix cf.
`N' for the previous filters). We include analysis for both filter
combinations.

\section{DATA PROCESSING TECHNIQUES}

All SCUBA data are reduced using the SURF (SCUBA User Reduction Facility)
package \citep{1998adass...7..216J,SURF} in conjunction with the Starlink
software environment \citep{SUN1}. To allow us to process the many thousands of
observations described in this paper we automated the reduction using the ORAC
data reduction pipeline (ORAC-DR)
\citep{1999adass...8...11E,1999adass...8..171J}.  The general reduction method
for SCUBA data is documented elsewhere \citep[e.g.,][]{SC11}. In this section
we comment briefly on the steps that we have tailored to our automated
procedure, or which require some discussion in this context.

\subsection{Extinction correction}

Once the data have been demodulated, corrected for beam switching and
flatfielded a critical calibration step is to correct for the atmostpheric
attenuation.  The difficulties of calculating the atmospheric opacity at
submillimetre wavelengths are well documented \citep[e.g.,][]{sr94} but
recent progress has been presented by \citet{ENA2002} who use skydips
taken at both the JCMT and the Caltech Submillimeter Observatory (CSO) to
provide a reliable record of the opacity spanning a period of several years. We
refer the reader to \citet{ENA2002} for a detailed discussion but give
a brief outline below.

Automated reduction relies critically on the quality of the opacity
measurement -- simply applying the nearest skydip or CSO tau measurement to
the data is dangerous for a number of reasons. Specifically, SCUBA skydips are
usually not taken frequently enough to be useful in variable conditions. In
addition, although the CSO tau measurements are taken much more frequently
(every 10 mins), the instrumental noise on the CSO tau meter is quite high, it
produces erroneously large readings occasionally and operates at a fixed
azimuth. 

With these points in mind we proceeded as follows: a polynomial was fitted to
the re-scaled \citep{ENA2002} CSO tau data in regions where the SCUBA skydips
were in reasonable agreement; times when the CSO tau meter either did not
agree with the SCUBA value or when the signal was varying rapidly were not
included in these fits. This method ensured that the atmosphere seen by the
tau meter was the same as that seen by SCUBA. In practice, the pipeline
ignores data taken in periods when no fitted extinction value is available.
Whilst this means that no data are reduced when, for example, the CSO tau
meter is broken it is a necessary requirement of the automation process.
The error of the 225~GHz opacity from the fit is usually of order 0.005; 
if this error is assumed to be the dominant error in the extinction then 
for a source at airmass 1.5 this
contributes to the final result an error of approximately 3 per cent at
850-\mum\ and 20 per cent at 450-\mum.

\subsection{Despiking}

Occasionally, the time-series data contains spikes (for example from cosmic
rays) and these must be removed. In the simplest case this involves analysing
each bolometer in turn, removing data that are significantly different to the
mean. The complication for SCUBA mapping is that each bolometer sees different
parts of the target during the observation (``jiggling'' on and off a source)
such that a simple mean no longer provides valid statistics. For long
observations it is possible to compare data points that are from similar
points on the sky (effectively equivalent to subtracting source signal from
the data prior to despiking) but for short observations, such as those from
bright calibrators described below, there are simply too few data points for a
reliable detection of spikes when near the source in question. For this work,
we have compromised by performing the despiking for each bolometer
independently but with a large enough clipping level (5 sigma) that we can
detect the largest spikes without fear of clipping real source structure. This
approach does mean that we are still susceptible to small spikes in the data
but the overall effect on the integrated sum will be minimal when averaged
over all the observations in this sample.

\subsection{Sky-noise removal}

Nodding and chopping do not remove all the atmospheric emission noise from the
data so an additional step is required that takes into account the fact that
the source signal is constant in time and the sky-noise is varying in time but
coherent across the array. Removal of sky-noise from jiggle maps requires
knowledge of regions of the map that are uncontaminated by source flux. Our
criterion of selecting only compact sources for analysis simplifies this
procedure enormously.  For pointing observations it is safe to use the median
of all the bolometers to determine the sky-noise signal; the pointings use an
azimuthal 60 arcsec chop-throw so that each map contains the source at the
centre of the map and two negative images with half the flux. The latter
images are present because the chop throw is smaller than the field-of-view
with the result that the negative beams of some of the `sky' bolometers fall
onto the central source.  For standard mapping observations the chop throw is
generally chosen to be larger than the field-of-view in order to avoid this
effect. It was, however, found that median sky removal was equally well
applicable to these observations. See \citet*{1998SPIE.3357..548J} and
\citet*{ENA2002} for examples of the sky-noise removal algorithm.

\subsection{Residual sky removal}

The sky-noise removal algorithm currently assumes that the sky signal adds a
d.c.\ offset across the whole array. In some extreme cases this is not true
and a linear gradient is visible across the array. To compensate for this
a plane is fitted to the regridded image and subtracted before the data
are processed. For the data considered here this procedure has only a minor
effect because the source is in the centre of the image but is performed for
completeness (when the sky-noise is a simple offset this step will have no
effect).

\subsection{Signal determination}

\label{sect:sigd}

The penultimate step in the reduction process is to determine the signal from
the source.  In the present case there is always a single point-like source at
the centre of the map and this simplifies the analysis.

Having first made sure that we have accurately located the centroid position
there are two ways to determine the signal from a source: (1) measure the peak
flux, and (2) measure the flux in an aperture.  Both of these approaches are
valid, we calculated both, and compare them in Section \ref{sect:fcf}.

The first step for case (2) is to select an aperture of suitable diameter.
For the special case of pointing observations the aperture diameter can be no
larger than 60~arcsec otherwise contamination from the negative images of the
central source becomes a problem. In general, there are two competing
considerations, the effects of which have to to traded-off to obtain the
optimum aperture size for the particular problem. On the one hand, it is
desirable to use as large an aperture as possible so that most of the time
dependent error beam pattern is enclosed for both source and calibrator. This
is especially important at 450-\mum\ where the error lobe contribution
typically exceeds 100 per cent for chop throws larger than about 2 arcmin. On
the other hand, increasing the aperture also increases the measured noise. The
final trade-off is thus between accuracy of calibration and signal-to-noise.
Experimenting with various apertures we found that 40~arcsec is a good choice
for point-like sources.  Increasing the aperture to 60~arcsec results in an
increase in signal of only 5--6 per cent at 850-\mum\ under most conditions
considered, and correspondingly 13 per cent at 450-\mum\ whereas the
signal-to-noise degrades by a factor of 2/3 (the error on the integrated sum
is proportional to the radius of the aperture). Apertures of significantly
less than 40~arcsec produced unacceptable scatter in the signals measured for
Uranus with the standard deviation increasing by almost a factor of 2 (see
Table~1).

The signal-to-noise is determined by looking at small regions towards the edge
of the map and determining the standard deviation of the pixels in the
aperture. Bolometers with uncharacteristically high noise are often apparent
when maps are made of faint sources. In order to avoid these and other effects
like spikes, which are not always easy to remove from small datasets, a number
of apertures are used and the one with the smallest standard deviation is
selected to determine the overall error in the integrated sum (the aperture
measures the noise-per-pixel).  We verified that this procedure is valid by
reducing a number of maps by hand.

Once the data have been reduced automatically there are inevitably a small
number of points that must be removed by hand because of problems with the
data acquisition, or with the instrument or telescope.  The main problems are
occasional pointing observations where the source is not centred on the array,
where the dish has been heated by the Sun and is grossly out of shape, when
SCUBA is not operating correctly, or when the polarimeter is attached.
For these cases, the observing logs and final data products are
inspected and the point is removed if warranted by the circumstances.

\begin{figure*}
\includegraphics[angle=-90,width=\textwidth]{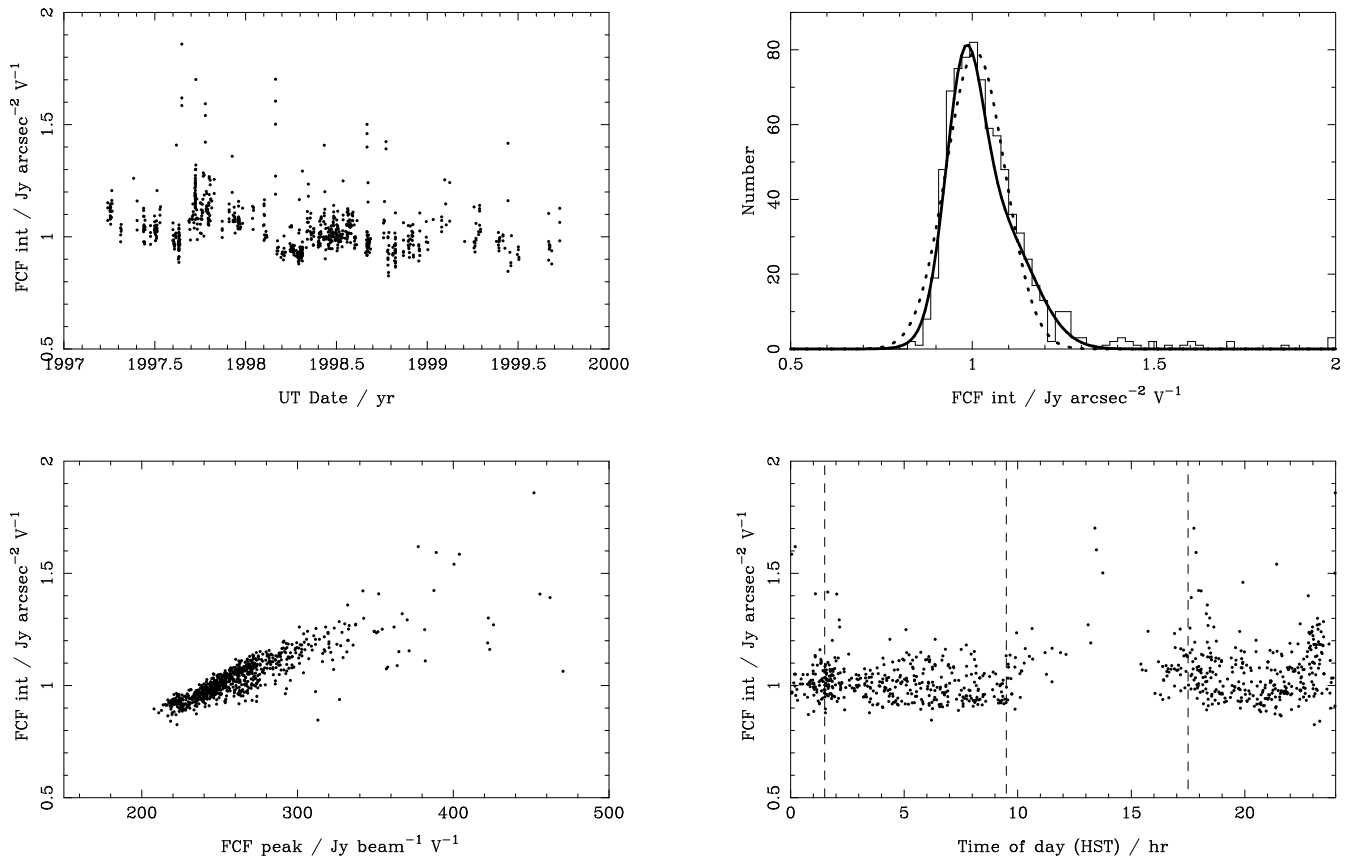}
\caption{Flux conversion factors derived for the 850N filter for Uranus. `FCF
  int' is the integrated flux conversion factor and `FCF peak' is the flux
  conversion factor derived from the peak of the source. The vertical dashed
  lines on the bottom right panel delineate the observing shifts of the JCMT.}
\label{fig:850N}
\end{figure*}
\section{Results}

\subsection{Flux conversion factor}
\label{sect:fcf}

\begin{figure*}
\includegraphics[angle=-90,width=\textwidth]{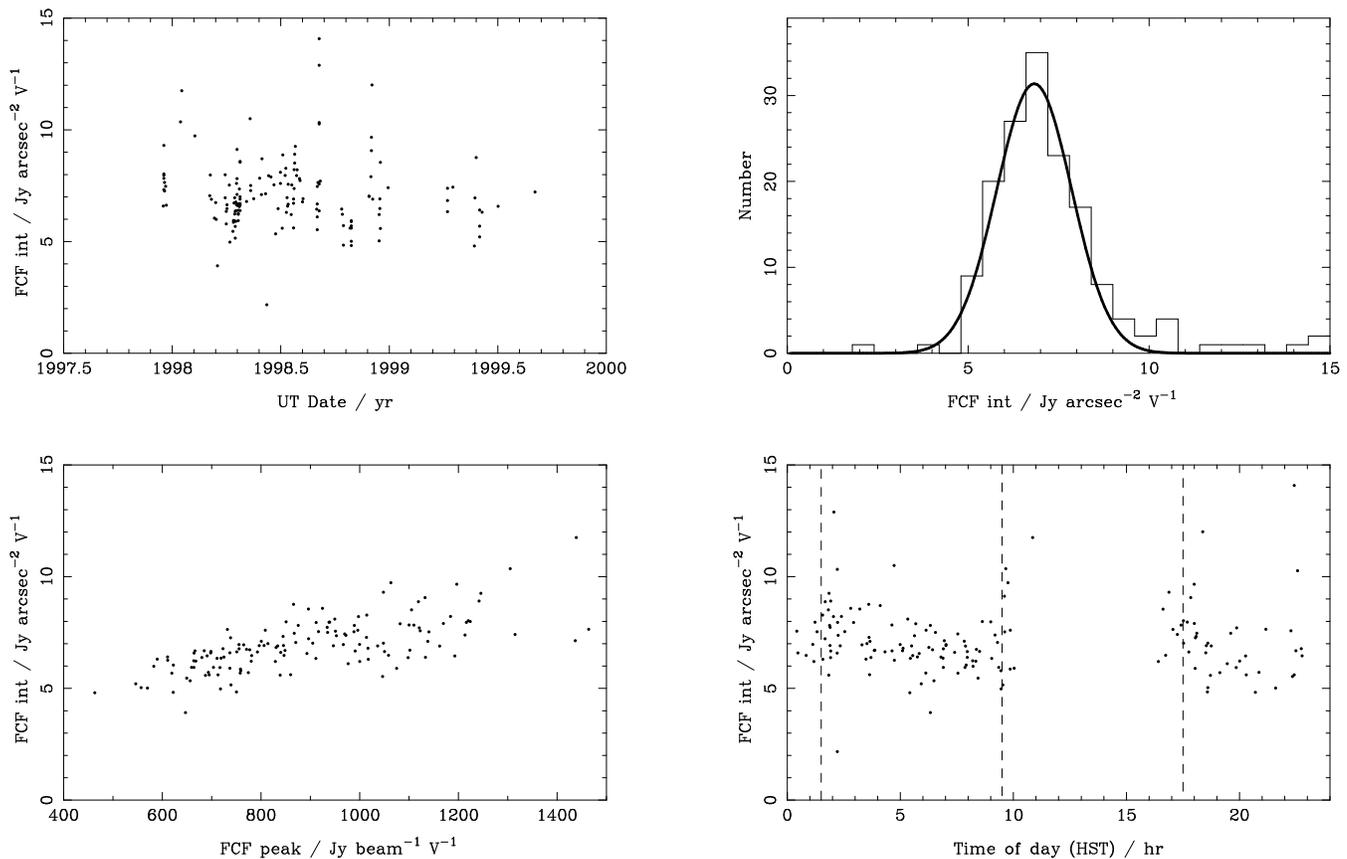}
\caption{Same as Fig.~\ref{fig:850N} but for the 450N filter.} 
\label{fig:450N}
\end{figure*}

The final step in the reduction process is the application of a suitable `flux
conversion factor' (FCF) to the reduced data so that the measured signal in
volts is converted into physically useful units such as Janskys. In order to
automate the reduction of arbitrary data we need to know the FCF history of
the telescope and instrument. We do this by measuring the response of our
primary calibrators (Mars and Uranus) throughout the period of interest and
comparing this with the expected flux derived from a standard model for
planetary fluxes \citep*{FLUXES,1993Icar..105..537G,1976ApJ...210..250W}.  The
FCF can take two forms: as mentioned in Section~\ref{sect:sigd} we can either
calibrate on the `peak' of a point source giving the traditional radio
astronomy units of Jy~beam$^{-1}$ ($FCF_{peak} = S_{peak}/V_{peak}$
where $S_{peak}$ is the flux in the beam and $V_{peak}$ is the measured peak
voltage), or we can `integrate' in an aperture giving units of Jy~arcsec$^{-2}$
($FCF_{int} = S_{tot}/[V_{int}A]$ 
where $S_{tot}$ is the total flux of the calibrator, $V_{int}$ is the
integrated sum in volts and $A$ is the pixel area in arcsec$^2$).
In this section we discuss the merits of the two schemes by applying the
automated pipeline reduction to Uranus.

The results of our Uranus reduction for the 850N and 450N filters are shown in
Figs.\ \ref{fig:850N} and \ref{fig:450N} respectively. Similar trends are seen
for the 450W/850W filter pair, and the FCF values are summarized in
Table~\ref{tab:fcf}.  The top left panel of Fig.\ \ref{fig:850N} shows the FCF
calculated by summing the signal in a 40 arcsec aperture plotted against time
for a period of $2-3$ yrs. In general the FCF is remarkably stable given the
variable weather/telescope/instrument parameters over this relatively long
time period. There is evidence for the FCF to be slightly higher during 1997
and early 1998 than in subsequent times. It is presently unclear as to why
this should be the case (if real) but we note that SCUBA was warmed up in
early 1998 because of varying base temperature problems and a similar, small,
change in FCF was seen following a warm up in the middle of 2000. Also
apparent from this panel is that a small fraction of the FCF values are
discrepant by a large amount ($> 50$ per cent). The effect is shown clearly in
the top right panel of Fig.\ \ref{fig:850N} where we show a histogram of the
integrated FCF values.  The dotted line shows a Gaussian fit to the data
giving a standard deviation of around 10 per cent as used by Robson et al.
(2001).  The solid line shows the result of fitting a double Gaussian to the
data (Table~\ref{tab:fcf} lists the average for the filter over the entire
period) which gives an improved fit. Note the tail of high FCF values.

The lower left panel of Fig.\ \ref{fig:850N} shows the `integrated' FCF versus
`peak' FCF.  Variations in the sensitivity of SCUBA connected with warm-ups
must contribute significantly to this general trend shown in this plot.
However, on any given night we believe that the linear trend, and hence the
dominant source of uncertainty in the calibration of SCUBA data is produced by
the thermal state of the antenna; it is well known that the focus of the
antenna changes rapidly around sunset and sunrise reflecting its changing shape
as it thermally relaxes. During these periods the observed beam pattern is also
variable because power is removed from the main beam and spread out into the
error lobes. The fact that we see a trend demonstrates that the error beam
spreads out beyond the $40$~arcsec aperture, but the key point to note is that
whereas the `peak' FCF varies by about a factor of two, the `integrated' FCF
varies by $< 50$ per cent.  Furthermore, the size of the chop throw can vary
the `peak' FCF by as much as $5-10$ per cent (the beam smears out a small amount
in the direction of the chop throw and the sensitivity is reduced) but affects
the `integrated' FCF much less.  The best reduction method for our automated
approach is thus to use the `integrated' FCF for flux determination.  A
corollary is that photometry mode observations, which are made with a single
pixel and a $2$ arcsec, $3\times$3 square jiggle pattern, and hence are calibrated
in an analogous manner to the FCF `peak' method described above, can at times
be susceptible to large calibration uncertainties. A quantitative assessment
would require detailed analysis of calibration data from individual nights
which is beyond the scope of the automated approach described in this paper,
but if high values of the FCF are measured then it may well be good observing
practice to use SCUBA in `jiggle-map' mode rather than photometry
mode. Although mapping observations are intrinsically less efficient than
photometry observations, the ability to measure the signal in a suitably sized
aperture will allow observations to be calibrated accurately during the period
when the antenna is changing shape.

The extreme values plotted in the lower left panel of Fig.~2 probably occur
when the antenna is grossly out of shape which may reflect the nature of the
observing, for example if observing was carried out in the daytime or if the
dish was pointed towards the Sun in early evening. These data points are
calculated from maps in which the beam shape is so obviously poor that they are
often removed by hand in the post-pipeline phase of the data reduction.
Evidence that this is indeed the case is provided in the bottom right panel of
Fig.\ \ref{fig:850N} where we plot `integrated' FCF versus time of day Here we
see that the majority of high FCF values ($> 1.3$) occur during the first half
of the night or after sunrise. Note also that the FCF can remain high until the
middle of the night in some cases but is almost always lower than about 1.2
during stable night-time conditions after midnight. This may indicate that once
the antenna has been heated it can take several hours to recover its optimum
shape.  Similar trends are seen at 450-\mum\ (Fig.\ \ref{fig:450N}) where a
single Gaussian provides a good fit the `integrated' FCF values giving a
standard deviation of about 20 per cent.  Experience shows that the
uncertainties derived from our automated approach are similar to those applied
by observers when processing their data `by hand', demonstrating the validity
of the method for point-like sources.

The ratio of `peak' to `integrated' FCF provides some idea of the magnitude of
the error beam within the aperture at these wavelengths. At 850-\mum\ the
ratio for a $40$ arcsec aperture is approximately 250 arcsec$^2$~beam$^{-1}$, an
area equivalent to a $15$ arcsec Gaussian beam (the measured value is
approximately 14.5 arcsec). At 450-\mum\ the ratio is approximately $120$
arcsec$^2$~beam$^{-1}$, equivalent to a $10$ arcsec beam (the main beam is
approximately $7.5$ arcsec).  The average error lobe contribution within a $40$
arcsec aperture is thus about $4$ per cent at 850-\mum\ and $40$ per cent at
450-\mum.

\begin{table}
\caption{Flux conversion factors determined from Uranus for three different
  apertures. There is evidence for a decrease in FCF for 850N in February
  1998 (a decrease of approximately 8 per cent). This table lists the average 
  for the 850N filter. The Peak FCF is only listed once for each filter since 
  the value is not dependent on aperture. All results are from a combination
  of chop throws between 60 and 120 arcsec.}.
\begin{tabular}{ccrclrcl}
Filter & Aperture & \multicolumn{3}{c}{Integrated} & \multicolumn{3}{c}{Peak} \\
       & diameter &\multicolumn{3}{c}{Jy arcsec$^{-2}$V$^{-1}$} &
       \multicolumn{3}{c}{Jy beam$^{-1}$V$^{-1}$}\\ 

& / arcsec & &&&&& \\ \hline
         
850N   & 20 & 1.43 &$\pm$&0.19 & 253 &$\pm$& 31\\
       & 40 & 1.00 &$\pm$&0.11 &     &     &  \\
       & 60 & 0.94 &$\pm$&0.10 &     &     &  \\
850W   & 20 & 1.19 &$\pm$&0.11 & 207 &$\pm$& 13 \\
       & 40 & 0.84 &$\pm$&0.04 &     &     &    \\
       & 60 & 0.80 &$\pm$&0.06 &     &     &    \\
       &    &      &     &     &     &     &\\
450N   & 20 & 10.0&$\pm$&2.6   & 855 &$\pm$& 260 \\
       & 40 & 6.9&$\pm$&1.5 &     &     & \\
       & 60 & 6.1 &$\pm$&1.2  &     &     & \\
450W   & 20 & 4.0&$\pm$&0.8 & 340 &$\pm$& 81  \\
       & 40 & 3.2&$\pm$&0.5 &     &     &   \\
       & 60 & 2.8&$\pm$&0.5 &  & &   \\
\hline
\end{tabular}
\label{tab:fcf}
\end{table}

\subsection{Mars}

\begin{figure}
\includegraphics[angle=-90,width=\columnwidth]{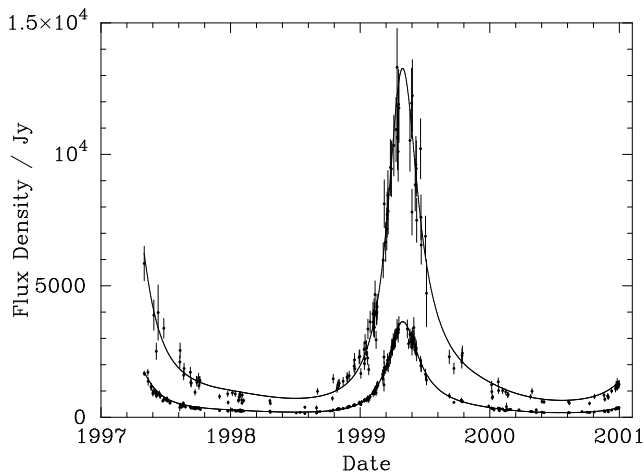}
\caption{Flux density measured for Mars at 850- (lower curve) and 450-\mum\
  (upper curve). The solid curve indicates the expected flux using the model
  of Wright (1976). Errors are derived by combining observations taken on the
  same night.}
\label{fig:mars}
\end{figure}

In order to test the output from the automated reduction, including a
test of the validity of our calibration, we tested the complete system
with observations of Mars. An aperture of 60 arcsec was used to
maximize signal (signal-to-noise is not an issue) and to guarantee the
inclusion of the majority of the flux (Mars had a maximum diameter of
15 arcsec during the period).  The results are shown in Fig.\
\ref{fig:mars} along with the expected flux provided by the
\textsc{fluxes} software \citep{FLUXES} which uses the model for mean
surface temperature derived by \citet{1976ApJ...210..250W} and
knowledge of the SCUBA filters. The excellent agreement between the
observed and predicted fluxes provides convincing support for our
automated reduction procedure and the Mars model.

\subsection{Secondary calibrators\label{sect:seccal}}

\begin{table}
\caption{Coordinates of the calibrator sources as defined in the JCMT pointing
  catalogue. All coordinates are J2000.}
\label{tab:seccalcoords}
\begin{tabular}{lll}
  & R.A. & Dec \\
\hline
HL Tau        &  04 31 38.4   & $+$ 18 13 59.   \\
CRL618        &  04 42 53.60 & $+$ 36 06 53.7 \\
OH231.8+4.2   &  07 42 16.94 & $-$ 14 42 49.1 \\
IRC+10216     &  09 47 57.38 & $+$ 13 16 43.7 \\
CRL2688       &  21 02 18.81 & $+$ 36 41 37.7 \\
\hline
\end{tabular}
\end{table}

Now that the system has been shown to generate reasonable results we can apply
these techniques to archive data. Here we present the results from processing
of the most common JCMT secondary calibrators (see Table
\ref{tab:seccalcoords}); the results from extragalactic radio pointing sources
are presented in \citet{RSJ2001}. Table \ref{tab:seccal} shows the results of
the automated reduction of the selected calibrator sources using a 40 arcsec
aperture. These data are selected in the same way as for Uranus with only
pointing and mapping observations retrieved from the data archive. This list
does not include potential candidates, such as VY~CMa and TW~Hya, mentioned in
\citet{1994MNRAS.271...75S} simply because the archive does not currently
contain enough observations to justify reliable automated
reduction.\footnote{By the end of the year 2000 there were only 12 suitable
  observations of VY~CMa and 20 suitable observations of TW~Hya in the
  archive. Of those only 2 of the VY~Cma and 6 of the TW~Hya observations had
  450-\mum\ data. At 850-\mum\ the results from automated reduction were
  2.1$\pm$0.1~Jy for VY~Cma and 1.46$\pm$0.2~Jy for TW~Hya} Since all these
sources are used for pointing there are significantly more observations
available at 850-\mum\ than at 450-\mum. IRAS~16293$-$2422 has not been
included in this analysis because of the extended nature of the source
\citep[e.g.,][]{2001clim.confE..93S}.

\begin{table}
\caption{Secondary calibrator fluxes for a 40 arcsec aperture. The fluxes in
  italics are from \citet{2001clim.confE..93S} and are shown for comparison
  even though they are derived from the peak and will therefore not include
  extended flux. The results for IRC+10216 and OH231.8 
  are the average values; see   \S\ref{sect:seccal} for more details on the
  variability of these
  sources. Fluxes are determined by fitting a Gaussian to the histogram of raw
  data to determine the peak and standard deviation. The number of observations
  is  given in brackets.}
\label{tab:seccal}
\begin{tabular}{lr@{~}c@{~}l@{~}cr@{~}c@{~}l@{~}c}
        & \multicolumn{4}{c}{850~$\mu$m} &
          \multicolumn{4}{c}{450~$\mu$m} \\ \hline
CRL618    &    4.69&$\pm$&    0.37 &(1445)&    12.1&$\pm$&    2.2 &(398)\\
          &\it 4.55&$\pm$&\it 0.20 &&\it 11.5&$\pm$&\it 1.5 &\\ 
          &        &     &         &&        &     &        &\\
CRL2688   &    6.39&$\pm$&    0.51 &(223)&    30.9&$\pm$&    3.8 &(57)\\
          &\it 5.9 &$\pm$&\it 0.20 &&\it 24&$\pm$&\it 2 &\\
          &        &     &         &&        &    &         &\\
OH231.8+4.2 &  2.75&$\pm$&    0.44 &(272)&    12.7&$\pm$&    2.2 &(59)\\
          &\it 2.46&$\pm$&\it 0.20 &&\it 10.5&$\pm$&\it 1.2&\\
          &        &     &         &&        &     &        &\\
HLTau     &    2.36&$\pm$&    0.24 &(287)&     9.9&$\pm$&    2.0 &(83)\\
          &\it 2.3&$\pm$&\it 0.15 &&\it  10.0&$\pm$&\it 1.2 &\\
          &        &     &         &&        &     &        &\\
IRC+10216 &    8.8 &$\pm$&     1.1 &(736)&    17.5&$\pm$&    4.5 &(169)\\
          &\it 6.1&     &   0.2      &&\it 13.0&     & 1.8       &\\
\hline
\end{tabular}
\end{table}

Table \ref{tab:seccal} indicates that the automated reduction shows excellent
agreement with \citet{2001clim.confE..93S}, especially at 850-\mum. The
fundamental difference is that the numbers presented here are integrated over
an aperture and will therefore also include flux from extended emission
whereas \citet{2001clim.confE..93S} calibrated his data in Jy/beam.
\subsubsection{HL Tau}

\begin{figure}
\includegraphics[angle=-90,width=82mm]{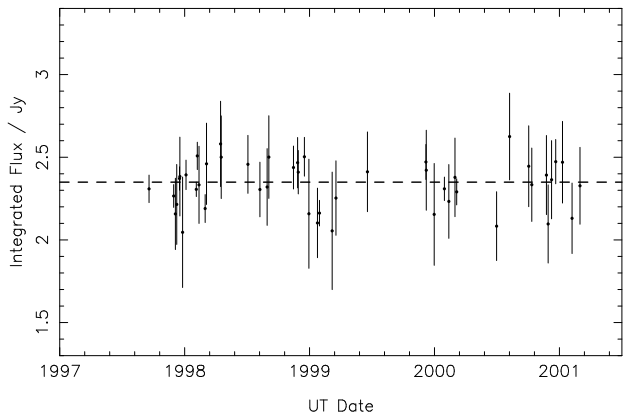}
\caption{Light curve for HL Tau at 850-\mum. Errors are derived
  by averaging observations over 5 day periods. The mean flux density is
  indicated by dashed lines.}
  \label{fig:hltau}
\end{figure}

The flux density shows excellent agreement with \citet{2001clim.confE..93S} at
both wavelengths indicating that there is no evidence for extended structure
out to $20$ arcsec radius when performing chopped observations.  Fig.\ 
\ref{fig:hltau} shows the light-curve. Whilst the plot raises the possibility
of source variation (possibly with a maximum at 1998.5, amplitude of 100~mJy
and period of $600$ d) a chi-square fit indicates that the data are
consistent with zero variability.
\subsubsection{CRL~618}

\begin{figure}
\includegraphics[angle=-90,width=82mm]{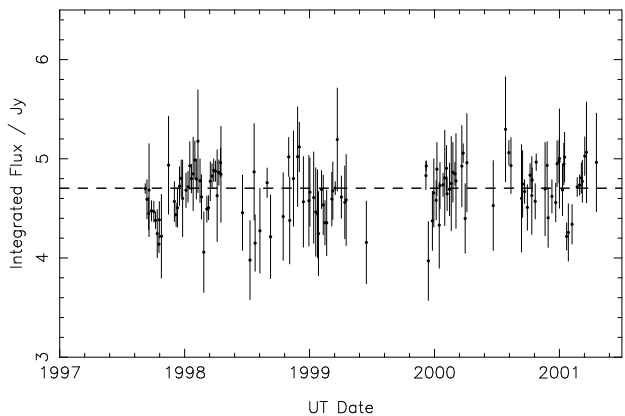}
\caption{Same as Fig.\ \ref{fig:hltau} but for CRL~618}
  \label{fig:crl618}
\end{figure}

CRL~618 is the most popular secondary calibrator in use at the JCMT. The flux
density shows excellent agreeement with \citet{2001clim.confE..93S} and
\citet{1994MNRAS.271...75S} at both wavelengths (the latter taken using UKT14
\citep{1990MNRAS.243..126D}) indicating that there is no evidence for extended
structure out to $20$ arcsec radius when performing chopped observations.
\citet*{1993ApJS...88..173K} indicate a possibility of source variability in
the millimetre region but Fig.\ \ref{fig:crl618} shows that no obvious
variability can be seen at 850-\mum\ between mid-1997 and mid-2001. Apparent
short term trends (of the order of $0.2$~yr) in the data stream are most likely
connected with the observing/reduction methodology or with FCF variations due
to dish heating rather than to source variability.
\subsubsection{OH~231.8+4.2}

We get excellent agreement with the results of \citet{2001clim.confE..93S}.
OH231.8 is a long period variable star and the time series show clear evidence
for variability at 850-\mum.  A least-squares fit of a sine-wave to the time
series indicates a period of $630 \pm 20$ d. This is somewhat lower, but
not necessarily inconsistent with, that seen in the near-infrared: $650$ d
\citep{1983MNRAS.203.1207F}, $684 \pm 40$ \citep{1984ApJ...276..646B} and $708
\pm 6$ \citep{1992ApJ...398..552K}, but our data are relatively noisy.
Indeed, \citet{1992ApJ...398..552K} suggest that the period may have changed
slightly over the years and it is therefore possible that the period has
shortened recently.

\begin{figure}
\includegraphics[angle=-90,width=82mm]{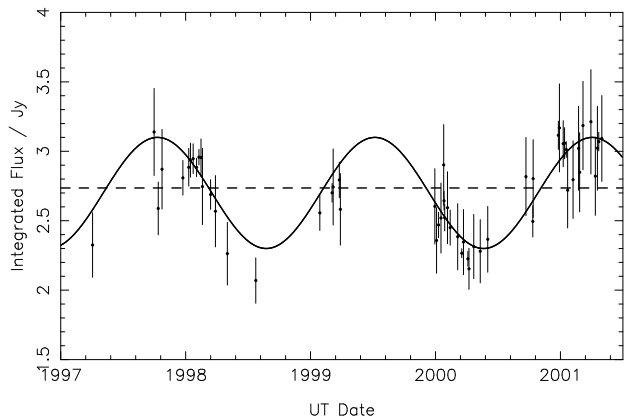}
\caption{Light curve for OH231.8 at 850-\mum. Errors are derived
  by averaging observations over 5 day periods. The mean flux density is
  indicated by dashed lines. A sine wave of period 630 days is shown.}
  \label{fig:oh231}
\end{figure}

Fig.\ \ref{fig:oh231} includes a sine-wave with a period of $630$ d of the
form:

\begin{equation}
S (Jy) = 2.75 + 0.4 \sin\left(2\pi\frac{Y - 1999.08}{1.75}\right)
\end{equation}
where $Y$ is the year, or alternatively:
\begin{equation}
S (Jy) = 2.75 + 0.4 \sin\left(2\pi\frac{JD - 2451207}{630}\right)
\end{equation}
where $JD$ is the Julian day.  If the sine wave is subtracted from the time
series data the error in the residual is reduced to $0.2$~Jy. The uncertainty in
the period makes it difficult to extrapolate the light curve from earlier
studies with any accuracy. Therefore it is not possible to reliably determine
the lag time between the light maximum in the near-infrared and that of the
submillimetre. The 450-\mum\ data are too noisy to show source
variability.

\subsubsection{IRC+10216 (CW Leo)\label{sect:irc10}}

This is a variable star with a period of $635$~d as seen in the optical
\citep{1989IBVS.3315....1A} and at 1.1-mm \citep{1994MNRAS.271...75S}.  When
compared to \citet{2001clim.confE..93S} and \citet{1994MNRAS.271...75S} this
source is clearly extended in our aperture. Using a source size of $8.9$~arcsec
at 850-\mum\ and $4$~arcsec at 450-\mum\ \citep{SandellWWW} the results from
\citet{2001clim.confE..93S} would be equivalent to 8.4- and 16.4-Jy in
excellent agreement with our results.

\begin{figure}
\includegraphics[angle=-90,width=82mm]{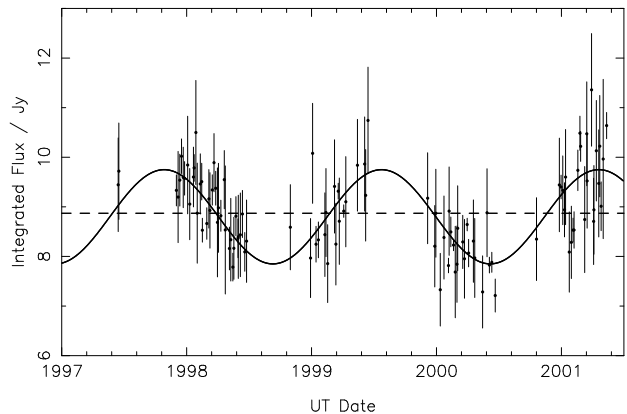}
\caption{Light curve for IRC+10216 at 850-\mum. Errors are derived
  by averaging observations over 5 day periods. The mean flux density is
  indicated by a dashed line. The sine wave has the same periodicity as the
  variation observed in the optical.}
  \label{fig:irc10}
\end{figure}

Fig.\ \ref{fig:irc10} shows the measured light-curve at 850-\mum\ along with a
sine-wave of period $635$~d of the following functional form:

\begin{equation}
S (Jy) = 8.8 + 0.95 \sin\left(2\pi\frac{Y - 1999.125}{1.74}\right)
\end{equation}
where $Y$ is the year, or alternatively:
\begin{equation}
S (Jy) = 8.8 + 0.95 \sin\left(2\pi\frac{JD - 2451224}{635}\right)
\end{equation}
where $JD$ is the Julian day.  If the sine-wave is subtracted from the
time-series data the error in the residual is reduced to $0.56$~Jy. These data
indicate that the source is at a maximum approximately $120$~d after the
maximum expected in the optical from \citet{1989IBVS.3315....1A} and $72$~d
($0.2$~yr) after the maximum expected from the 1.1-mm observations of
\citet{1994MNRAS.271...75S}. It is likely that the discrepancy in phase
between the 850-\mum\ and 1.1-mm data is not significant given the difficulty
in obtaining the 1.1-mm data with a single pixel instrument. The 450-\mum\
data are too noisy and do not show reliable evidence of variability.
\subsubsection{CRL~2688}

\begin{figure}
\includegraphics[angle=-90,width=82mm]{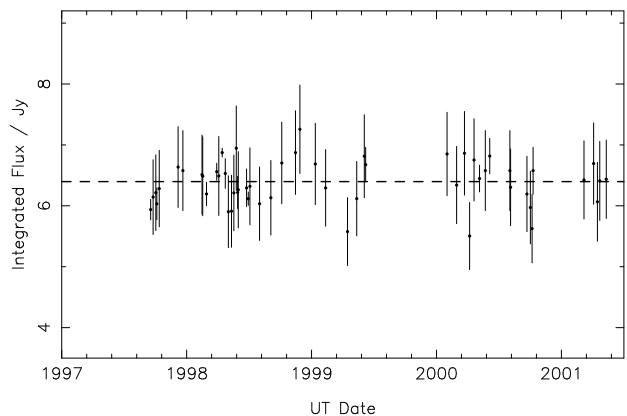}
\caption{Same as Fig.\ \ref{fig:hltau} but for CRL~2688}
  \label{fig:crl2688}
\end{figure}

CRL~2688 has a $25$ per cent higher 450-\mum\ flux here than presented in
\citet{2001clim.confE..93S} suggesting that the source is extended at this
wavelength. A map was generated combining the data from a number of stable
nights from the archive to test this, taking particular care that any pointing
shifts are accounted for during the reduction.  The measured `source size' of
$9.5$~arcsec is larger than that measured from CRL~618 (known to be
unresolved), using the same techniques, of $8.0$~arcsec \citep[see
also,][]{IMC2001}. This corresponds to a deconvolved source size of
approximately $5$~arcsec and would account completely for the lower flux
measured by \citet{2001clim.confE..93S}.  With a source size of 5~arcsec we
would expect a much larger discrepancy at 450-\mum\ than at 850-\mum\ which is
indeed the case. The time-series in Fig.\ \ref{fig:crl2688} indicates that the
flux density has been constant since the start of 1998.
\section{CONCLUSIONS}

This paper presents a means of automating the reduction and calibration of 850-
and 450-\mum\ SCUBA data. The limitations of the technique are investigated by
analysing archival data for the standard submillimetre calibration source,
Uranus. We show that our method gives excellent agreement between observed and
predicted submillimetre fluxes for Mars. It provides a useful tool for
investigating trends in instrument and telescope parameters -- these studies
are essentially made practical for the first time with our automated procedure,
and will ultimately lead to a better understanding of the uncertainties
involved in the calibration of data from the JCMT and other submillimetre 
observatories.

We use our automated procedure to calculate light-curves and integrated fluxes
for selected, compact JCMT secondary calibrators. The periodic variability of
IRC+10216 is confirmed and there is strong evidence to support similar
behaviour for OH231.8 at 850-\mum; the first time this variability has been
confirmed in the submillimetre. The remaining calibrators have been
constant during the period 1997.5 to 2001.5.

\section*{ACKNOWLEDGMENTS}

The James Clerk Maxwell Telescope is operated by the Joint Astronomy Centre in
Hilo, Hawaii on behalf of the parent organizations PPARC in the United
Kingdom, the National Research Council of Canada and The Netherlands
Organization for Scientific Research. J.A.S. acknowledges the support of a
PPARC PDRF.  We acknowledge the support software provided by the Starlink
Project which is run by CCLRC on behalf of PPARC.  Thanks to Wayne Holland,
Remo Tilanus, Iain Coulson and G\"{o}ran Sandell for useful discussions on
this topic.


\bsp

\end{document}